%% file: main.tex
\documentclass{article}

\usepackage[utf8]{inputenc}
\usepackage{authblk}
\usepackage{setspace}
\usepackage[margin=1.in]{geometry}
\usepackage{graphicx}
\usepackage{subcaption}
\usepackage{amsmath}
\usepackage[per-mode=repeated-symbol]{siunitx}
\usepackage{lineno}
\usepackage{xcolor}
\definecolor{mylilas}{RGB}{170,55,241}

\usepackage[style=nejm, 
citestyle=numeric-comp,
sorting=none]{biblatex}
\title{Compact, Intense Attosecond Sources Driven by Hollow Gaussian Beams}

\author[1,2*]{Rodrigo Martín-Hernández}
\author[3*]{Melvin Redon}
\author[3]{Ann-Kathrin Raab}
\author[3]{Saga Westerberg}
\author[3]{Victor Koltalo}
\author[3]{Chen Guo}
\author[3]{Anne-Lise Viotti}
\author[1,2]{Luis Plaja}
\author[1,2]{Julio San Román}
\author[3]{Anne L'Huillier}
\author[3]{Cord L Arnold}
\author[1,2]{Carlos Hernández-García}

\date{}
\affil[1]{Grupo de Investigación en Aplicaciones del Láser y Fotónica, Departamento de Física Aplicada, Universidad de Salamanca, Salamanca, Spain.}
\affil[2]{Unidad de Excelencia en Luz y Materia Estructuradas (LUMES), Universidad de Salamanca, Salamanca, Spain.}
\affil[3]{Department of Physics, Lund University, P.O. Box 118, 22100, Lund, Sweden}

\begin{document}

\maketitle

\begin{abstract}
High-order harmonic generation (HHG) enables the up-conversion of intense infrared or visible femtosecond laser pulses into extreme-ultraviolet attosecond pulses.
However, the highly nonlinear nature of the process results in low conversion efficiency, which can be a limitation for applications requiring substantial pulse energy, such as nonlinear attosecond time-resolved spectroscopy or single-shot diffractive imaging.
Refocusing of the attosecond pulses is also essential to achieve a high intensity, but difficult in practice due to strong chromatic aberrations.
In this work, we address both the generation and the refocusing of attosecond pulses by sculpting the driving beam into a ring-shaped intensity profile with no spatial phase variations, referred to as a Hollow Gaussian beam (HGB).
Our experimental and theoretical results reveal that HGBs efficiently redistribute the driving laser energy in the focus, where the harmonics are generated on a ring with low divergence, which furthermore decreases with increasing order.
\textcolor{black}{Although generated as a ring,} the attosecond pulses can be refocused with greatly reduced chromatic spread, therefore reaching higher intensity. 
\textcolor{black}{This approach enhances the intensity of refocused attosecond pulses and enables significantly higher energy to be delivered in the driving beam without altering the focusing conditions. These combined advantages open pathways for compact, powerful, tabletop, laser-driven attosecond light sources.}
\end{abstract}

\input{intro_suggestion}

\input{section_1.tex}

\input{Discussion}
\printbibliography
\end{document}


\maketitle

\begin{abstract}

In this Supplementary Information we first analyze the role of aberrations in the Hollow Gaussian Beam (HGB) in High Harmonic Generation (HHG). This allows us to understand subtle differences between theoretical and experimental results in the main text. Second, we study the role of the intrinsic dipole phase of HHG in the refocusing of attosecond pulses.
\end{abstract}

\section{Effect of aberrations in the driving Hollow Gaussian Beam in High Harmonic Generation}

\begin{figure}[t!]
    \centering
    \includegraphics[width=0.9\linewidth]{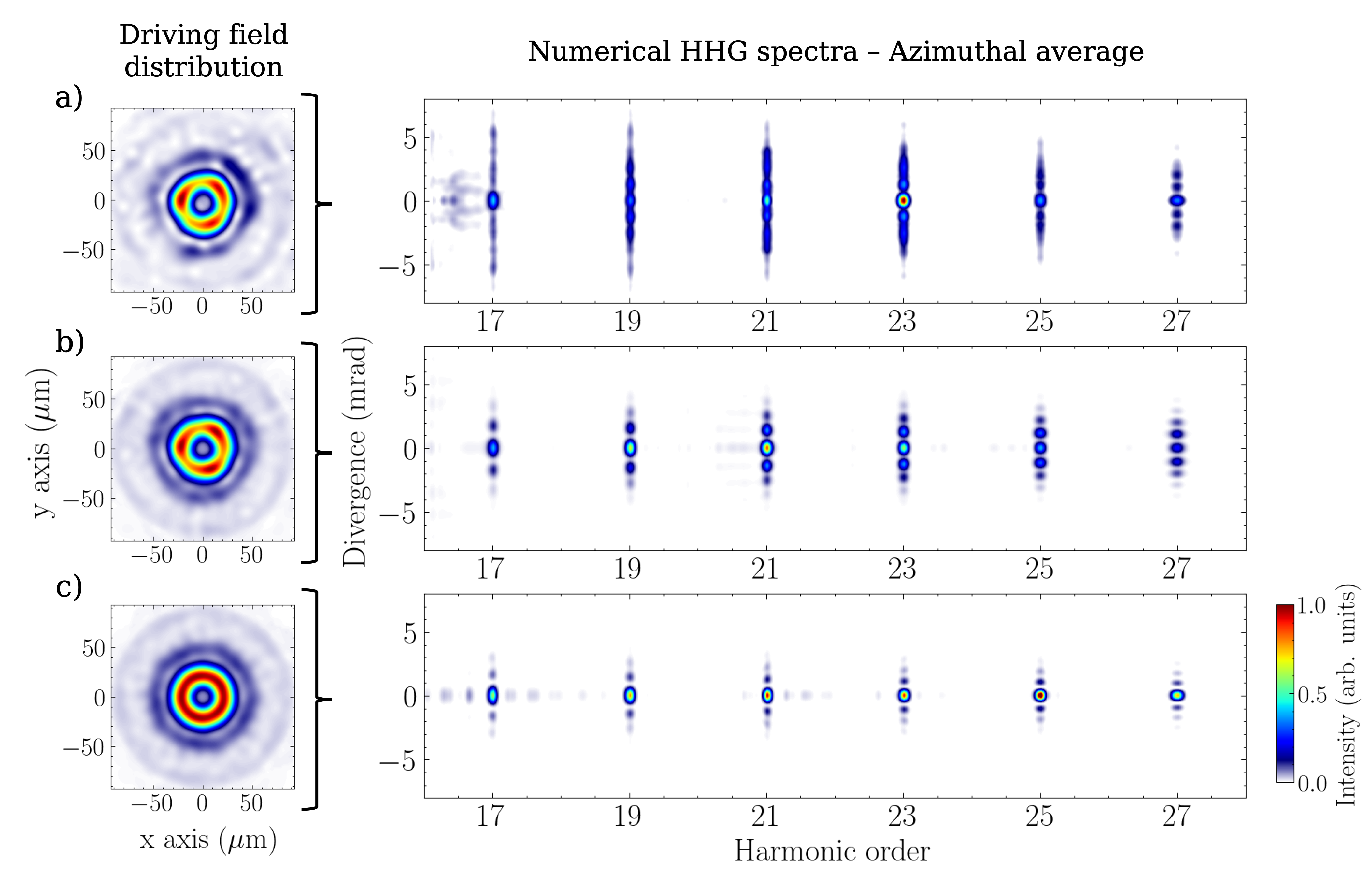}
    \caption{Effect of the HGB driving field homogeneity in far-field HHG emission obtained trough numerical simulations. The left column shows the driving field distribution (input of the HHG simulations) and the right column depicts the HHG spectral distribution from the macroscopic simulations. Three driving field distributions are considered, from experimentally abberated, to perfect HGB: a) the experimentally retrieved driving field ; b) three $120^\circ$-rotated copies of the experimental field, and c) five $72^\circ$-rotated copies. The driving field parameters are the same as for Fig. 3 in main text.}
    \label{fig:experimentalHHGSim}
\end{figure}
The high coherence of HHG makes this process highly sensitive to the quality of the driving field, where aberrations and distortions of the driver beam can substantially affect the harmonic emission properties. When comparing the spatial profile of the high-order harmonics driven by HGB in Figs. 2 b) (experimental) and 3 c) (theoretical), we qualitatively observe the decreasing divergence trend  with harmonic order. However, the theoretical intensity profiles exhibit a particular feature that is not present in the experimental results: the harmonics present a central intense spot surrounded by secondary rings.

This discrepancy can be explained by the aberrations present in the experimental driving field. Indeed, the spatial driving field distribution shown in Fig. 1 b) in the main text exhibits subtle intensity and phase modulations that are not present in the perfect HGB assumed in theory (Fig. 3 c). Here we perform simulations considering the experimental HGB profiles as an input to our simulations to analyze the effect of such imperfections.  

In Fig.  \ref{fig:experimentalHHGSim} a) we show the simulated far-field harmonic spectra driven by the experimentally retrieved HGB driving field. Note that in contrast to the results presented in Fig. 3 c) of the main text where a perfect theoretical HGB was assumed, here the harmonic profiles are averaged along the azimuthal coordinate. In this situation, the macroscopic HHG numerical simulations predict a harmonic emission with similar features to the experimental observations, where the outer rings in each harmonic are faded out. This indicated that the driving field aberrations inhibit the appearance of the secondary outer rings. In order to corroborate this hypothesis, in Figs. \ref{fig:experimentalHHGSim} b) and c) we provide theoretical simulations considering smoother aberrations in the HGB driving field. We have crafted symmetrized driving beams as superposition of rotated copies of the experimentally retrieved field profile shown in a). In particular, in panel b) we have overlapped three copies of the experimental field, each one rotated by $120^\circ$, while in panel c) we have superimposed five copies rotated by $72^\circ$. We observe how the simulated harmonic emission start to exhibit the central spot plus the secondary rings as the driving field distribution is more symmetric. In the latter case, the harmonic far-field distribution is almost indistinguishable from the perfect HGB case, Fig. 3 c) in the main text.

\begin{figure}[t]
    \centering
    \includegraphics[width=0.5\linewidth]{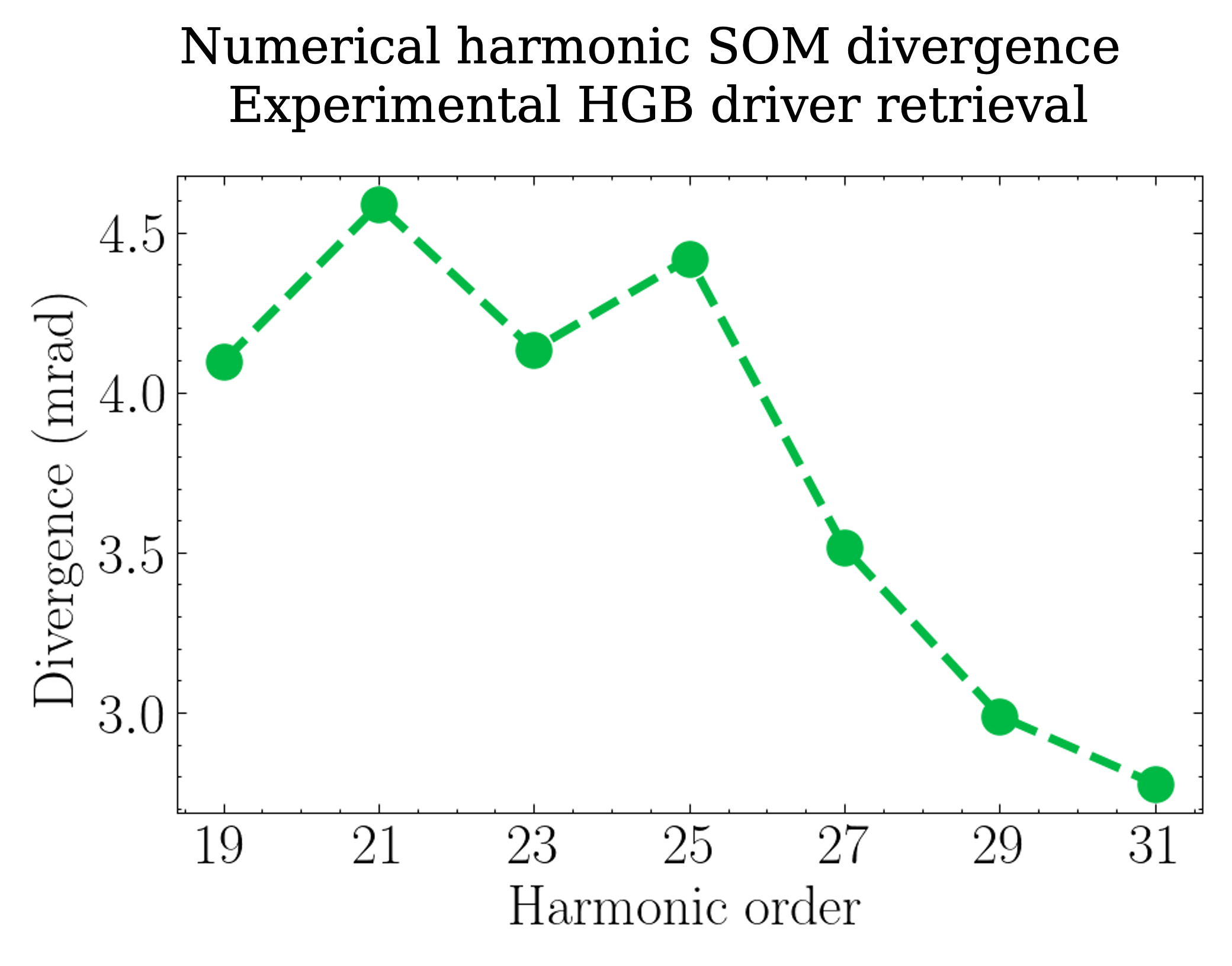}
    \caption{Harmonic SOM divergence from the numerical HHG simulations in Fig. S1 a), assuming the experimental HGB driving field retrieval. The predicted harmonic divergence values are similar to the measured values in Fig. 2 c) for the HGB case.}
    \label{fig:divergence_exp}
\end{figure}

In Fig. \ref{fig:divergence_exp} we show the far-field SOM divergence obtained for the HHG numerical simulations in Fig. \ref{fig:experimentalHHGSim} a), where the experimental driver is assumed. In this case, we still observe the decreasing harmonic divergence trend and we are able to predict divergence values closer to the experimental measurements in Fig. 2 c).

\section{Role of the dipole phase on the attosecond refocusing}
In the main text we have already analyzed the role of intrinsic dipole phase over the harmonic far-field divergence, whose effect is more pronounced for the Gaussian driving field. Here we study the role of the intrinsic dipole phase on the synthesis of the attosecond pulse train after refocusing the EUV emission. We confirm that the origin of the spatiotemporal coupling observed for the attosecond pulse trains after refocusing---depicted in Fig. 4 c)---is mainly due to the dipole phase.

\begin{figure}[t]
    \centering
    \includegraphics[width=0.9\linewidth]{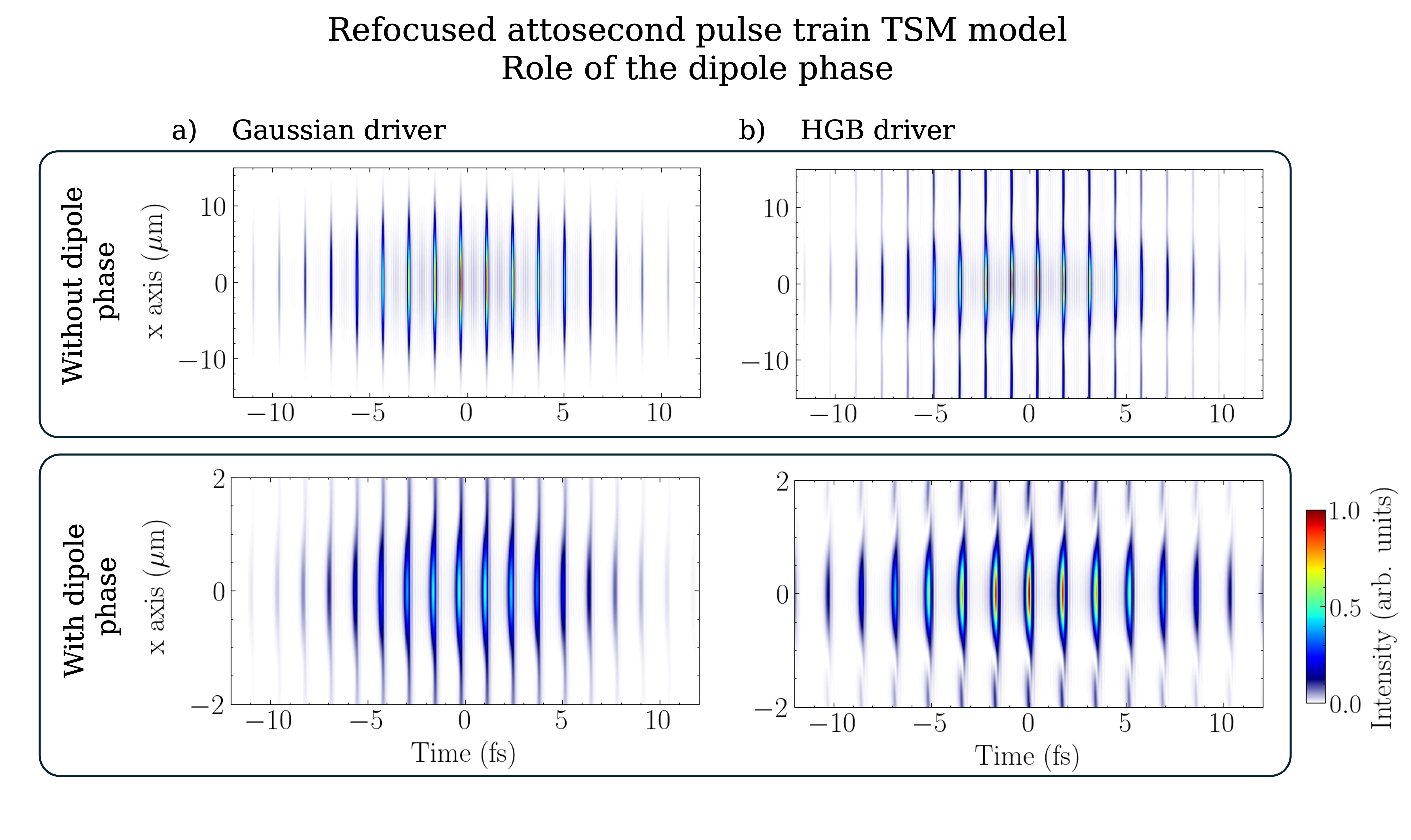}
    \caption{Role of the intrinsic dipole phase over the refocused attosecond pulse train. The attosecond pulse train obtained via refocusing the HHG emission overlaping the harmonic orders $25^\textrm{th}$-$49^\textrm{th}$, using the TSM model for: a) the Gaussian driver and b) the HGB driver. The top row corresponds the refocused attosecond pulse train from the TSM without the dipole phase, while the bottom row shows the results for the TSM accounting for the dipole phase. Note that the spatial extent of the attosecond pulse train in the top row is $\sim 5$ times larger.}
    \label{fig:TSM_APT}
\end{figure}

To corroborate this hypothesis we have performed theoretical simulations through the simpler TSM model (see main text), where we can artificially neglect the dipole phase term. In Fig. \ref{fig:TSM_APT} we depict the attosecond pulse trains for the Gaussian (a) and HGB driving fields (b) neglecting (top) or including (bottom) the contribution from the intrinsic dipole phase \cite{wikmark2019spatiotemporal}. The HHG emission is assumed to propagate $40\:\mathrm{cm}$ from the generation plane, and is subsequently refocused using a focusing element with 20 cm focal length, matching the parameters of Fig. 4 in the main text. The attosecond pulse trains are calculated at the propagation distance where the maximum attosecond pulse intensity is obtained, as also performed in Fig. 4 of the main text. 

For the Gaussian driver case (a), we observe that only when the dipole phase is included we are able to recover the curved wavefront distribution for each individual attosecond pulse within the train. When the dipole phase is neglected, all attosecond pulses are emitted with a flat wavefront. For the HGB driving case (b), the behavior is similar. The spatiotemporal couplings arise from the intrinsic dipole phase. We thus conclude that the chromatic aberrations induced by the intrinsic dipole phase result in the spatiotemporal couplings observed after refocusing the far-field emission in HGB-driven HHG, similarly to the spatiotemporal couplings observed when focusing broadband IR/femtosecond pulses trough lenses with chromatic aberrations \cite{bor1992distortion}.

Moreover, for both driving fields, we see that if the dipole phase is not considered, the spatial extent of the attosecond pulse train is larger---see comparison between Fig. S3 b) and Fig. 4 c). This indicates that the harmonics are generated from a smaller, virtual source due to the contribution of the dipole phase \cite{wikmark2019spatiotemporal}, that upon refocusing  the virtual source is imaged.

\printbibliography

%% file: Intro_suggestion.tex
\section{Introduction}


The generation of coherent extreme-ultraviolet (EUV) and soft x-ray radiation via high-order harmonic generation (HHG) has become crucial for advancing the understanding of ultrafast physical phenomena down to the attosecond time scale, enabling a broad range of scientific and technological applications. These include probing physical functions at the ultrafast time scales, such as thermal transport \cite{Siemens2010}, elastic scattering \cite{Karl2018}, many-body behavior \cite{Lee2024}, photoelectron states \cite{Laurell2025}, enabling high-resolution imaging and metrology \cite{Gardner2017}, or probing molecular dynamics \cite{Calegari2014, Severino2024} relevant to chemical reactions in biology, to mention some.

In the HHG process, an intense femtosecond (fs) infrared (IR) driving field is focused into a noble gas jet or gas cell. The underlying physics can be explained through the semiclassical three-step model \cite{Corkum1993, Schafer1993}. First, the target atom undergoes tunnel ionization in the high-intensity driving electric field. Next, the liberated electron is accelerated in the continuum. Finally, upon return to the parent ion, the electron may recombine, resulting in the emission of high-frequency radiation in the form of odd-order harmonics of the driving field. The resulting harmonics acquire an intensity- and order-dependent dipole phase that depends on the electron trajectory, imprinting a chirp to the attosecond pulses---known as \emph{attochirp}.

As a highly nonlinear process---reaching up to 5000 harmonic orders \cite{Popmintchev2012}---HHG suffers from low conversion efficiency on the order of $10^{-5}$ to $10^{-7}$ \cite{falcao-filho_analytic_2009}.
Particular applications however, such as nonlinear EUV spectroscopy\,\cite{Kretschmar2024SciAdv} or single-shot diffractive imaging\,\cite{MalmOPEX2020}, require high EUV/soft x-ray pulse energy and good refocusing properties in order to efficiently transport the radiation from the generation gas target to an experimental end station. 
Unfortunately increasing the intensity of the IR driving pulse does not lead to higher harmonic flux. At laser intensities above $\SI{1e15}{\watt\per\centi\meter\squared}$, the gas target enters the barrier suppression ionization regime, where the HHG conversion efficiency decreases drastically \cite{Moreno1995}. As an alternative to enhance the HHG flux, the spatial refocusing conditions can be carefully tailored. When HHG is driven with beams with Gaussian transverse profile, the intensity dependence of the dipole phase\,\cite{Lewestein1995} introduces chromatic aberrations by imprinting a harmonic dependent positive wavefront curvature.
As a result, the harmonic orders feature different divergences and appear to originate from different longitudinal positions along the optical axis\,\cite{wikmark2019spatiotemporal, Quintard2019}. 
Efficient re-focusing from the generation gas target to an experimental end station, where a small spot size and high intensity are required, while also maintaining the attosecond pulse duration, is therefore challenging.
Adjusting the driving beam wavefront by optimizing the relative position of the gas jet and beam focus can counteract the wavefront curvature induced by the harmonic dipole phase, enabling direct refocusing of specific harmonic orders \cite{wikmark2019spatiotemporal, Quintard2019}. Since the dipole phase is harmonic dependent, this optimization works over a limited spectral window, preventing the simultaneous refocusing of all harmonic orders and, consequently, the generation of higher-intensity attosecond pulses.
Driving beams that approximate a flat-top transverse intensity profile in the gas target, produced either via phase plates \cite{constant2012spatial} or spatial light modulators (SLM) \cite{treacher2020increasing}, have proven effective in reducing the divergence across all harmonic orders and enhancing EUV focusability by mitigating chromatic aberrations \cite{dubrouil2011controlling,veyrinas2023chromatic,Hoflund2021UltrafastScience}.
In this context, the emergence of structured driving beams---spatially tailored in their amplitude, phase and polarization properties---has opened exciting possibilities over the last decade \cite{2023_bliokh_Roadmap, 2023_shen_Roadmap}. Examples include the generation of high-orbital angular momentum (OAM) EUV optical vortices \cite{Hernandez-Garcia2013, Gariepy2014, Geneaux2016, Pandey2022}, EUV vector beams \cite{Hernandez-Garcia_17_vectorbeamsHHG}, temporal varying OAM pulses with self-torque \cite{Rego2019torque}, circularly-polarized attosecond pulses \cite{Huang2018}, attosecond vortex pulse trains \cite{delasHeras2024}, or EUV spatio-temporal optical vortices \cite{Martin-Hernandez2024}, among others.

In this work, we demonstrate that structured driving beams can be designed to enhance the refocused flux of \emph{all} harmonic orders, thereby increasing the intensity of attosecond pulses.
We suggest structuring the driving field into a ring-shaped intensity profile with no transverse phase variations---known as Hollow Gaussian Beam (HGB) \cite{Cai2003}.
Using a SLM, we generate an IR beam, which becomes a HGB at the position of an argon gas jet.
Our experimental and theoretical results show that, unlike standard Gaussian beams where the harmonic divergence increases with the harmonic order, the use of HGBs enables the emission of EUV high-order harmonics whose divergence decreases with the harmonic order.
Our theoretical simulations further reveal that, in contrast to Gaussian driving beams, the harmonic dipole phase plays a negligible role in the focusing properties of the harmonics. When using HGB driving beams, high-order harmonics are generated in an extended ring-shaped area with an efficiency comparable to that of Gaussian driving beams. 
By imaging the correct plane, the generated harmonics can be refocused into a small spot with little, order-independent, longitudinal spread leading to greatly reduced chromatic aberrations of the attosecond pulse.
As a result, we predict an increase in attosecond pulse intensity of a factor of up to three compared to optimized Gaussian driving field conditions. Our study presents a robust alternative to improve attosecond pulse focusing, enhancing their intensity for applications in metrology, imaging, x-ray diffraction and nonlinear x-ray spectroscopy, while opening the possibility to the use of high-power laser systems as bright attosecond pulse sources.

%% file: section_1.tex
\begin{figure}[t!]
    \centering
    \includegraphics[width=1\linewidth]{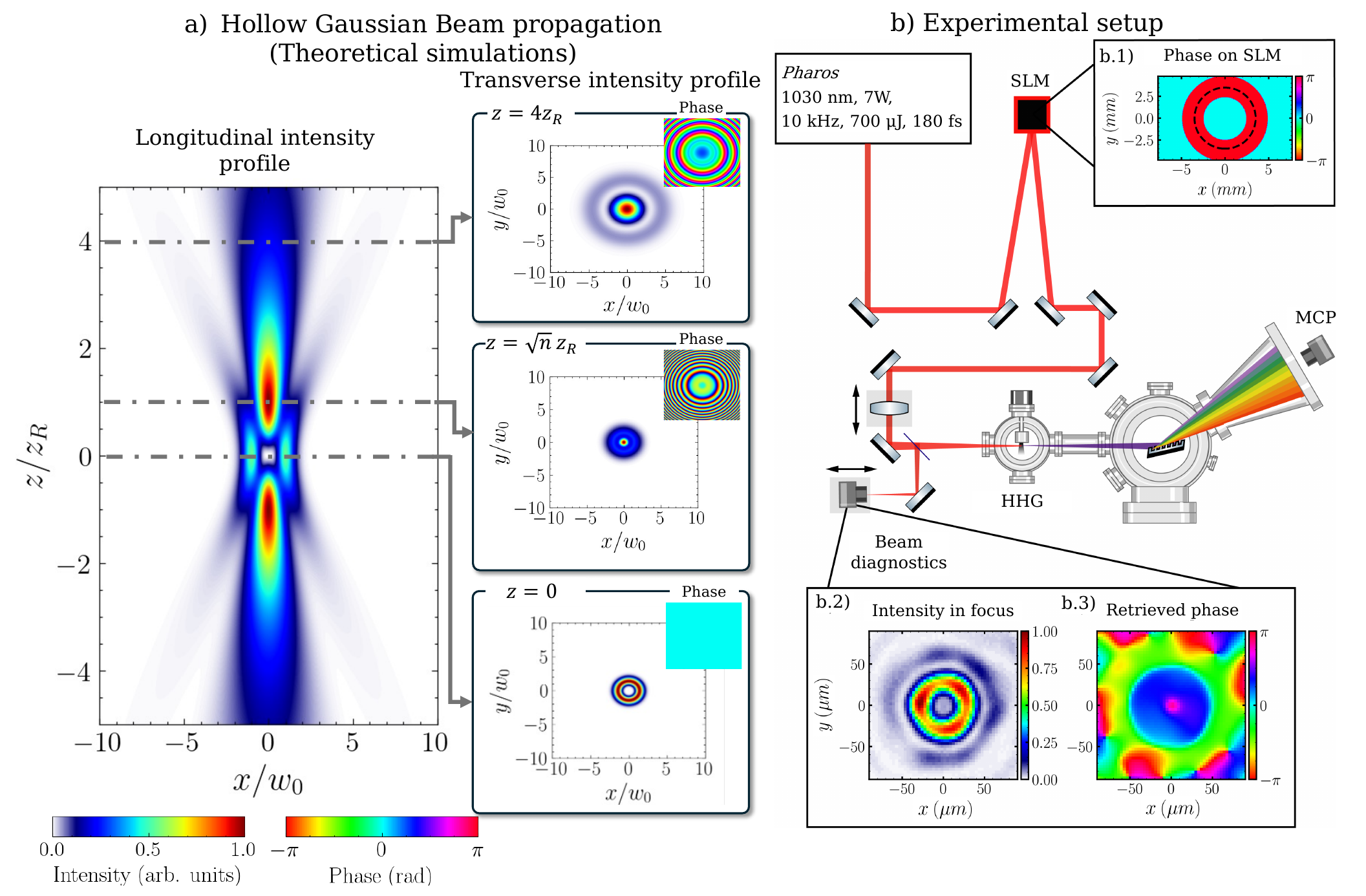}
    \caption{Hollow Gaussian beam propagation properties and generation to drive HHG. a) illustrates the intensity profile of a HGB with $n=1$ as a function of the reduced propagation distance $z/z_R$, extracted from the analytical expression Eq. (\ref{eq:HGB}). At the focal plane ($z=0$), the beam exhibits a ring-shaped intensity profile with no transverse phase variations. b) shows the experimental setup for driving HHG with a HGB. Panel b.1) shows the phase pattern applied to the SLM to \textcolor{black}{approximate} a HGB at the focal plane where an argon gas jet is placed. The dashed line represents $1/e^2$ of the input Gaussian beam. The resulting intensity profile and the retrieved phase at the focus, is shown in panels b.2) and b.3), respectively. The intensity plots are normalized and share the same color scale.}
    \label{fig:setup}
\end{figure}

\section{Results}

HGBs are not eigenmodes of free-space propagation, and are named after their structure at the focal plane, featuring a ring-shaped intensity profile and a flat transverse phase distribution. This contrasts with Laguerre-Gaussian beams, which are eigenmodes of propagation and can carry a twisted phase profile characterized by the topological charge. When using HGBs to drive HHG, it is instrumental to study their propagation dynamics to design an appropriate experimental setup. HGBs can be described as a polynomial expansion of fundamental Laguerre-Gaussian modes \cite{Cai2003}. In the paraxial approximation, considering propagation along the $z$-axis, the spatial dependence of the electric field in cylindrical coordinates $(\rho,z)$ is given by:
\begin{align}
\label{eq:HGB}
    E_n(\rho,z) = C_0\frac{n!}{2}\sum_{m=0}^n(-1)^m \binom{n}{m}\frac{(1+iz/z_R)^{m}}{(1-iz/z_R)^{m+1}}L_m\left(2\frac{\rho^2}{w^2(z)}\right)\nonumber \\ \times\exp{\left(-\frac{\rho^2}{w^2(z)}\right)}\exp{\left(-ik\left(z+\frac{\rho^2}{2 z\left(1+z_R^2/z^2\right)}\right)\right)},
\end{align}
where $n$ is the order of the HGB, $C_0$ is a normalization constant, $L_m$ is the $m^\mathrm{th}$ order Laguerre polynomial, and $k=2\pi/\lambda_0$ is the wavenumber related to the wavelength $\lambda_0$. The beam waist evolution along the propagation distance is defined by $w^2(z) = w^2_0\left(1+ z^2/z_R^2\right)$, where $w_0$ and $z_R=kw_0^2/2$ are the waist in the focal plane and the Rayleigh length, respectively. In the focal plane, $z=0$, Eq.~\eqref{eq:HGB} reduces to $E_n(\rho, 0) = C_0(\rho^2/w_0^2)^n e^{-\rho^2/w_0^2}$. Note that the complete spatiotemporal description of the field is given by 
$E_n(\rho,z, t) = E_n(\rho,z) f(t) e^{i\omega_0 t}$, where $f(t)$ defines the temporal envelope and $\omega_0$ is the pulse central frequency.

In Fig.~\ref{fig:setup} a), we present $|E_1(\rho,z)|^2$ along the propagation direction ($z/z_R$). Note that HGBs exhibit cylindrical symmetry around the propagation axis. In the focal plane ($z=0$), the characteristic ring-shaped intensity profile with a flat phase distribution is presented (see corresponding inset). As the beam propagates away from the focal plane ($|z|>z_R$), the annular intensity distribution gradually transforms into a quasi-Gaussian profile, surrounded by secondary rings (see inset for $z=4z_R$). In particular, the maximum intensity is not at the focal plane, but symmetrically before and after the focal plane at approximately $z=\pm \sqrt{n} z_R$.

\begin{figure}[t]
    \centering
    \includegraphics[width=0.8\linewidth]{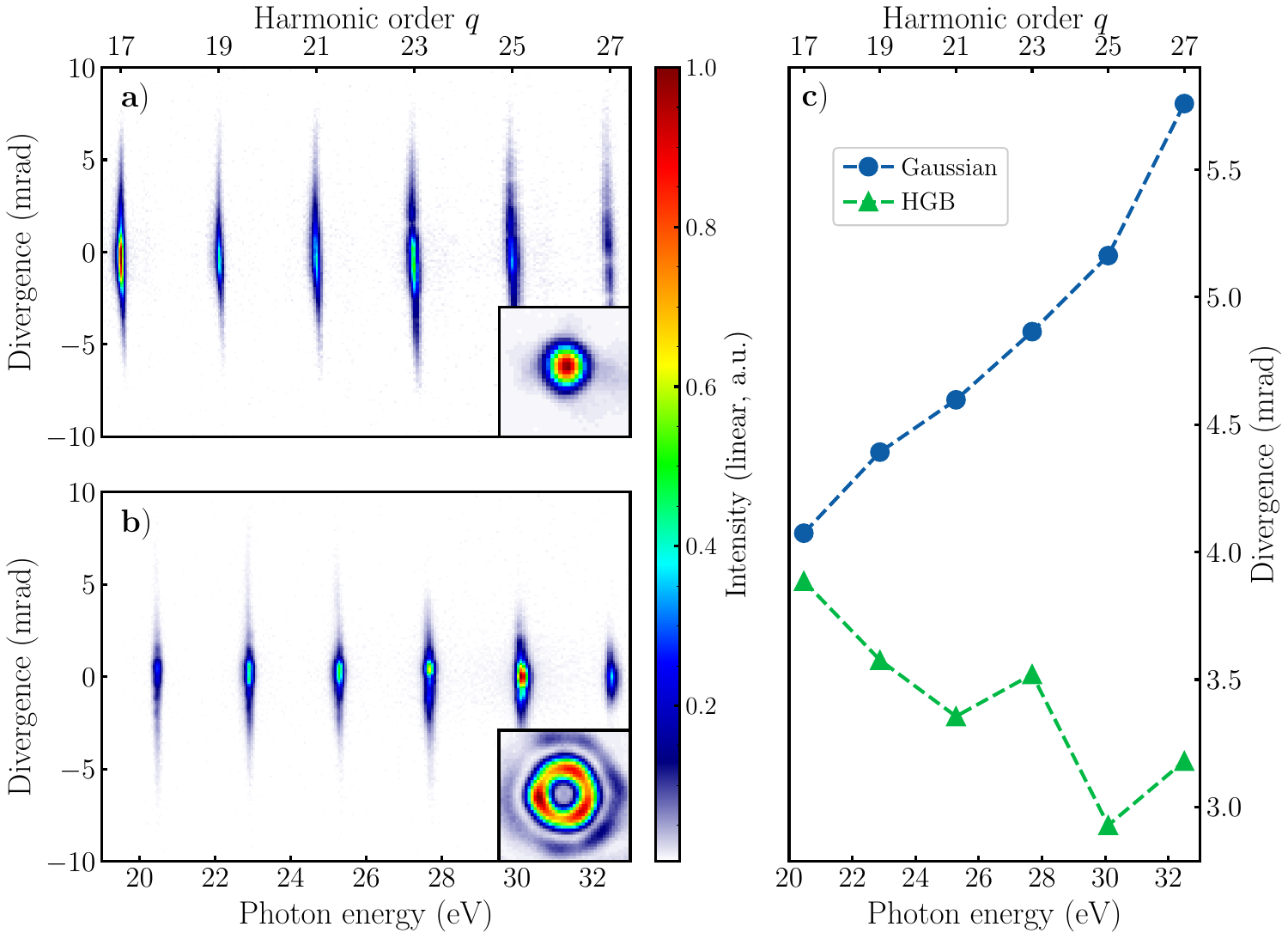}
    \caption{Comparison of experimental harmonic spectra and divergence when driving HHG with a Gaussian beam and a HGB. In both cases the Ar gas jet is placed at the focus position. Panels a) and b) show the spatially dependent harmonic spectra recorded in the far-field using a Gaussian and a HGB driving beam, respectively (the insets depict the measured driving field intensity profiles). Panel c) compares the far-field divergence, calculated using the second order moment for the Gaussian beam (blue) and the HGB 
 (green). As opposed to the Gaussian beam case, the far-field divergence of the harmonics generated by the HGB decreases with the harmonic order.}
    \label{fig:results_experiment}
\end{figure}

Figure ~\ref{fig:setup} b) presents the experimental setup used to study HHG driven by an HGB. An Ytterbium laser system (Pharos, Light Conversion) operating at a wavelength of $\SI{1030}{\nano\meter}$, with a pulse duration of $\SI{180}{\femto\second}$ full width at half maximum (FWHM), pulse energy of $\SI{700}{\micro\joule}$, and a repetition rate of $\SI{10}{\kilo\hertz}$, delivers $\SI{7}{\watt}$ of average power. The laser beam is directed on a SLM (Santec SLM300), which modulates the phase to shape the beam. In Fig.~\ref{fig:setup} b.1), the phase pattern applied to the SLM is displayed, consisting of a ring-shaped region with a $\pi$ phase jump, designed to \textcolor{black}{approximate} a HGB at focus, as shown in the intensity (Fig.~\ref{fig:setup} b.2) and phase (Fig.~\ref{fig:setup} b.3) profiles. After reflection on the SLM, the beam is focused by a lens with a focal length of $\SI{20}{\centi\meter}$, mounted on a motorized translation stage. The beam waist of the Gaussian beam is equal to $\SI{20}{\micro\meter}$. This corresponds to a Rayleigh length of $\SI{1.2}{\milli\meter}$. A portion of the beam is picked off just before the vacuum chamber and sent to a diagnostic camera, also mounted on a motorized translation stage. This setup allows the camera to capture the intensity profile in different planes along the focal region, enabling real-time control and adjustment of the beam profile. Since the SLM plane and the focal plane are related by a Fourier transform, the phase at the focal plane can be retrieved from intensity measurements. Specifically, the Gerchberg–Saxton retrieval algorithm enables reconstruction of the complex field at the focus by iteratively propagating between the SLM and the focal plane via Fourier and inverse Fourier transforms, while enforcing the measured intensity constraint at the focal plane and a known amplitude or phase constraint at the SLM \cite{gerchberg1972practical,fienup1982phase}. The retrieved phase is shown in Fig.~\ref{fig:setup} b.3). This provides an additional tool to refine and compensate for aberrations, optimizing the generation process \cite{raab2024highly}. The remaining portion of the beam enters the vacuum chamber, where HHG occurs in an argon gas jet from a nozzle with a $\SI{40}{\micro\meter}$ diameter opening and a backing pressure of $\SI{2}{\bar}$. The medium length is measured to be $\sim\SI{100}{\micro\meter}$, i.e. a tenth of the Rayleigh length. The high harmonic beam is then spectrally resolved using a curved EUV grating (600 lines/mm, radius of curvature = 5649 mm)  and detected by a phosphor screen coupled to a multi-channel plate (MCP).

The experimental far-field harmonic spectral distribution is shown in Fig.~\ref{fig:results_experiment}, comparing the results obtained with a Gaussian beam (a) and a HGB (b). In both cases, the gas jet is placed at the focus position. For a fair comparison between the Gaussian-driven and the HGB-driven cases, we kept a constant peak intensity of the Gaussian beam before the SLM. Consequently, although the overall energy is similar, the HGB distributes the energy over a larger effective area, resulting in a $4$-fold  peak intensity reduction. The spatial distribution of high-order harmonics from the 17th to the 27th is presented. Figure ~\ref{fig:results_experiment}c) shows the far-field harmonic divergence, calculated using the second order moment (SOM) for each harmonic order, for both the Gaussian (blue line) and HGB (green line) driving beams. While the harmonic spectrum generated by the Gaussian beam shows an increasing divergence with the harmonic orders, as expected from previous works \cite{wikmark2019spatiotemporal}, the trend is reversed for the HGB, where the divergence decreases as the harmonic order increases. Moreover, we demonstrate how the high-order harmonic spectra generated from HGB leads to improved refocusing properties, such as reduced chromatic dispersion, and the larger driving energy in the gas-jet results in more intense attosecond pulses, as discussed in the next section.



%% file: Discussion.tex
\section{Discussion}

To understand the experimental harmonic divergence behavior observed when using the HGB driver, we perform advanced HHG simulations that account for both microscopic (quantum wavepacket dynamics) and macroscopic (harmonic phase-matching and propagation) effects. Specifically, we compute the far-field harmonic emission by combining the Maxwell electromagnetic field propagator with the full-quantum strong field approximation \cite{Hernandez-Garcia2010}. This method has been successfully applied in several HHG experiments driven by structured light beams \cite{Pandey2022, Rego2019torque, Hernandez-Garcia_17_vectorbeamsHHG, Rego2022necklace, Huang2018, delasHeras2024, Martin-Hernandez2024}. We assumed an IR driving field with a wavelength of $\SI{1030}{\nano\meter}$, and a beam waist of $w_0 = \SI{15}{\micro\meter}$. The temporal envelope is defined as $f(t) = \sin^2\left(\pi t/t_{max}\right)$, where $t_{max}$ represents the full-width pulse duration. In the simulations $t_{max} = 32\:\textrm{optical cycles}$ (corresponding to $\SI{33} {\femto\second}$ FWHM). Due to computational constraints, the simulated pulse duration is shorter than that used in the experiment. The driving beam (defined by Eq.~\eqref{eq:HGB}) is focused into an infinitesimally thin argon gas jet, which is a good approximation for a medium length much shorter than the Rayleigh length, which is the case in our experiment.

\begin{figure}[t]
    \centering
    \includegraphics[width=1\linewidth]{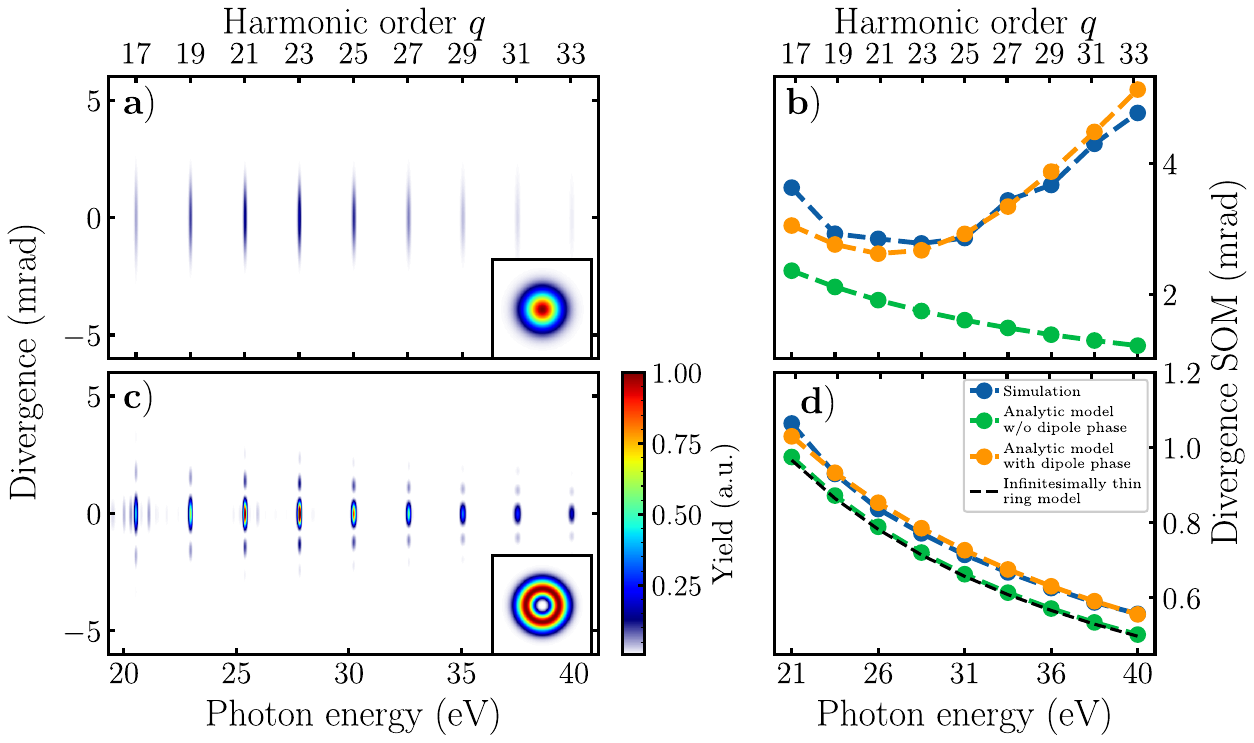}
    \caption{Comparison of the harmonic spectra and divergence for a Gaussian driver and a $n=1$ HGB driver obtained through numerical simulations. Panels a) and c) show the spatially resolved far-field high-order harmonic spectra for the Gaussian and the HGB driving fields, respectively (the insets depict the driving field intensity profiles). Panels b) and d) represent the far-field harmonic divergence (in blue) against the harmonic order, extracted from the numerical data in panels a) and c), respectively. In both cases, we compare the simulation results against those from the TSM including (orange) and neglecting (green) the harmonic dipole phase. For the HGB driving field case, additionally the divergence extension calculated with an infinitesimally thin ring model is presented (dashed black line).}
    \label{fig:theoretical_results}
\end{figure}

In Fig.~\ref{fig:theoretical_results} we present the simulated spatially resolved harmonic far-field spectra for Gaussian-driven a) and $n=1$ HGB-driven c) HHG. In the simulations, we maintained the same peak intensity $1.7\times10^{14}\:\text{W/cm}^2$ for the Gaussian beam and the HGB driving fields, translating into a pulse energy $\sim3.7$ times larger in the latter case. \textcolor{black}{It is important to note that in HGB-driven HHG, the harmonics exhibit a central, low-divergent spot surrounded by secondary rings. This intensity spatial distribution is expected from the far-field propagation of a HGB, as shown in Fig. 1 a) ($z=4z_R$). However, this prediction contrasts with the experimental HHG spatial distribution shown in Fig. 2 b), where the secondary rings are not visible. As discussed in the Supplementary Information S1, the absence of the outer rings is attributed to the aberrations of the experimental driving field}, \textcolor{black}{and the resolution and sensitivity of the MCP in the EUV spectrometer}

We quantify the divergence of the HHG radiation through the SOM of each harmonic profile, shown in Figs.~\ref{fig:theoretical_results} b) and Fig.~\ref{fig:theoretical_results} d) in blue. For the HGB driving field, the SOM is calculated over the central part of the harmonic beam, avoiding the secondary rings in each harmonic order.  The trends in harmonic beam divergence align with the experimental measurements in Fig. \ref{fig:results_experiment}, where the divergence increases with harmonic order for the Gaussian-driven case and decreases for the HGB-driven case.  \textcolor{black}{Overall, the absolute harmonic divergence for the theoretical HGB driver (neglecting the secondary rings) is significantly lower---$\sim 3$ times lower for the $21^{st}$ order and $\sim 8$ times lower for $33^{rd}$ order.}

To understand the underlying physics governing harmonic divergence in both cases, we employ the thin slab model (TSM), a well-established semi-analytical framework for macroscopic far-field harmonic beam propagation \cite{Hernandez-Garcia_2015}. In the TSM, the $q$-th harmonic near-field is approximated as $E_q^\mathrm{nf} = |E_0|^p e^{iq\phi_0 + i\phi_i}$, where $E_0$ and $\phi_0$ represent the spatial driving field and phase distributions of the fundamental driving field at an infinitesimally thin generation plane, $p\sim4$ is an effective non linear order that accounts for the HHG non-perturbative behavior and $\phi_i$ is the intensity-dependent harmonic dipole phase \cite{wikmark2019spatiotemporal}.

The green and orange curves in Figs.~\ref{fig:theoretical_results} b) and \ref{fig:theoretical_results} d) depict the divergence values predicted by the TSM in the far-field, including and neglecting the harmonic dipole phase, respectively. For HHG driven by a Gaussian beam (Fig.~\ref{fig:theoretical_results} c)), the TSM accurately predicts the increasing divergence trend only when the harmonic dipole phase term is considered. Since the harmonic dipole phase depends on the radially varying driving field intensity, it introduces a positive wavefront curvature in the harmonic beams, acting as a chromatic \textcolor{black}{defocusing} lens. Upon propagation, this wavefront curvature, combined with diffraction, results in increasing divergence, as already demonstrated in \cite{wikmark2019spatiotemporal}. In contrast, when HHG is driven by a HGB (Fig.~\ref{fig:theoretical_results} d)), the radial variation of the driving intensity profile is less pronounced, and as such the dipole phase has a negligible impact on harmonic divergence. Indeed, the harmonics are predominantly generated in the ring of maximum intensity, effectively washing-out the radial dependence of the dipole phase. As a result, the harmonics exhibit \textcolor{black}{reduced divergence} in the far-field, as consequence of the flat phase near-field harmonic profile. This behavior is corroborated by the dashed black curve in Fig.~\ref{fig:theoretical_results} d), which shows the TSM predictions when considering HHG only within the infinitesimally thin ring of maximum intensity, located at $\rho_{max}= \sqrt{n}w_0$. The Fraunhofer far-field harmonic emission $U_q(\theta)$ within this model is described analytically as $U_q(\theta) \sim 2\pi\rho_{max}J_0(2\pi\rho_{max}\theta/\lambda_q)$, where $\theta$ is the divergence angle and $J_0(\cdot)$ is the Bessel function of order zero. The SOM divergence for the central part, i.e. up to the first zero of the Bessel function, for $q$-th harmonic beam can be approximated to $\mathrm{SOM}_q\approx\lambda/ q\pi\rho_{max}$. 

 
\begin{figure}[t!]
    \centering
    \includegraphics[width=1\linewidth]{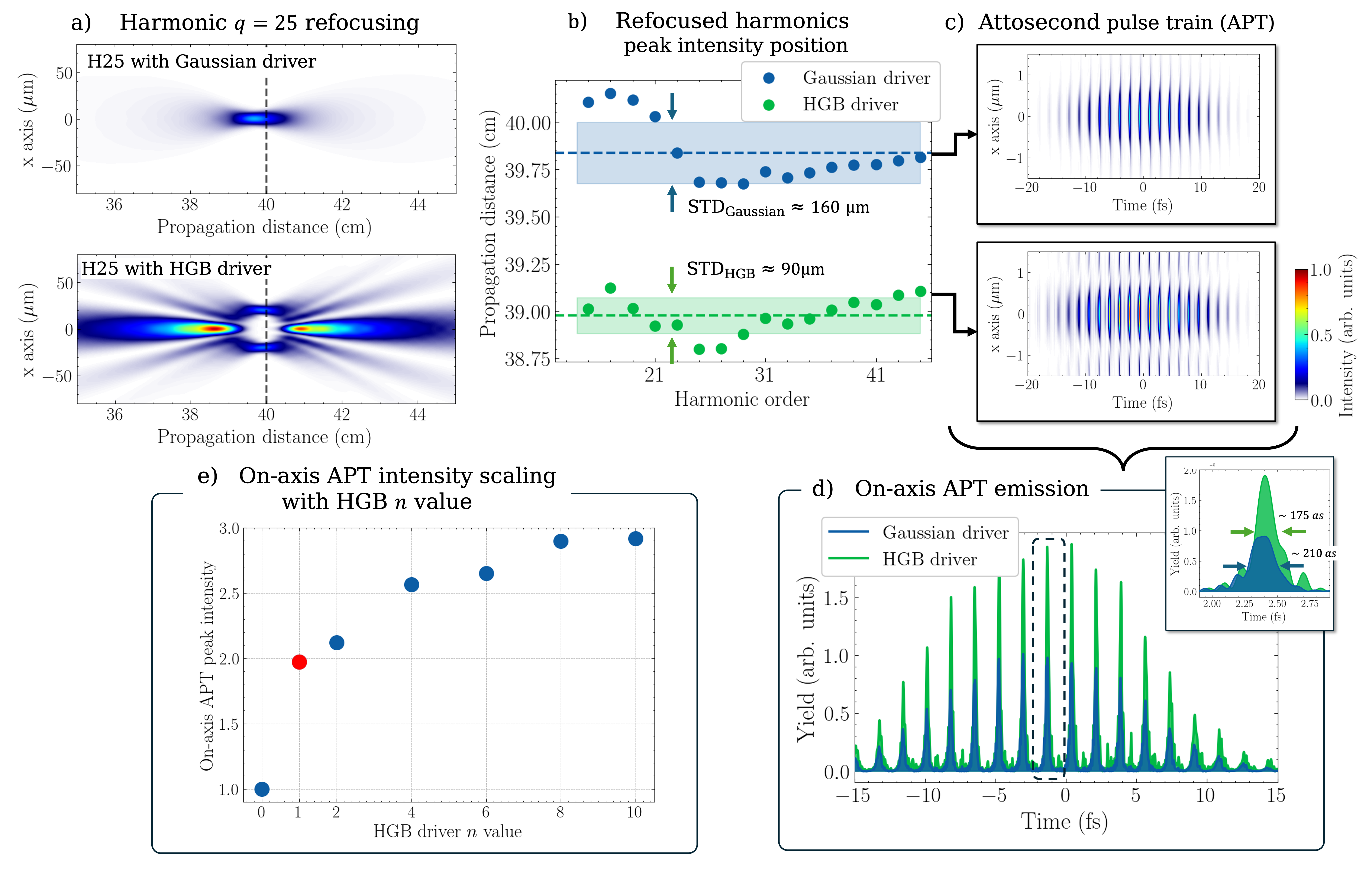}
    \caption{Simulated refocusing properties of high-order harmonics and attosecond pulses driven by \textcolor{black}{Gaussian and} HGBs. Panel a) illustrates the focusing dynamics for the 25$^{\text{th}}$ harmonic order generated with a Gaussian (top) and a $n=1$ HGB driver (bottom), after reflecting on a parabolic mirror with a $f=20\:\text{cm}$ focal length. In the latter case, the harmonic beam inherits the propagation dynamics of the IR HGB driving field. Noticeably, a hot-spot intensity region appears before the focal plane. Panel b) shows  the longitudinal position of the refocused harmonic peak intensity as a function of the harmonic order position for the Gaussian driver (blue) and the $n=1$ HGB driver (green). In both cases, the dashed lines show the average position ($z\approx 39.8\:\rm cm$ for the Gaussian driver and $z\approx 39\:\rm cm$ for the HGB driver) and the solid patch represent the standard deviation in the harmonic position (\textit{STD}). Panel c) depicts the spatiotemporal attosecond pulse train distributions obtained by filtering out the harmonic spectra in Fig. \ref{fig:theoretical_results} a) and c) below the $23\mathrm{rd}$ harmonic order. The attosecond pulse train is caluclated at the position given by the average harmonic peak intensity locations, indicated by the dashed lines in panel b). Panel d) corresponds with on axis attosecond pulse intensity profiles for the Gaussian (blue) and $n=1$ HGB (green) driving fields. The inset panel show the FWHM pulse durations. Panel e) show the refocused attosecond pulse train peak intensity scaling on-axis for higher-order HGBs, as a function of $n$, normalized to the peak intensity for the Gaussian-driven case ($n=0$). The red dot illustrates the case $n=1$ presented in panel d).}
    \label{fig:attosecond_pulses}
\end{figure}



The HGB-driven HHG translates the non-trivial HGB propagation dynamics to the high-order harmonics with a significant impact on the refocusing properties of the associated attosecond pulses.
To illustrate this effect, we first show in Fig.~\ref{fig:attosecond_pulses} a) the simulated refocusing dynamics of the 25$^{\text{th}}$ harmonic, after refocusing in a $f=\SI{20}{\centi\meter}$ parabolic mirror.  
In the Gaussian-driven case, the propagation dynamics of the 25$^{\text{th}}$ harmonic closely resembles those of a Gaussian beam. Notably, the peak intensity position (i.e. the harmonic beam focus) does not align with the focal plane of the parabolic mirror. This mismatch arises from the wavefront curvature induced by the dipole phase. Since the dipole phase is harmonic-dependent, the focal position varies significantly across different harmonic orders, as previously demonstrated in \cite{wikmark2019spatiotemporal}. This behavior is illustrated by the blue dots in Fig.~\ref{fig:attosecond_pulses} b), which represent the focal positions of the Gaussian-driven harmonic beams along the propagation axis ($z$) in our numerical simulations. The standard deviation of the harmonic focal positions is found to be $\rm{STD_{Gaussian}}\approx\SI{160}{\micro\meter}$. \textcolor{black}{Note  as a reference that the estimated Rayleigh length of the 25$^\textrm{th}$ harmonic order is $4.8\:\textrm{mm}$ (see Fig. 4 a)).} 

In the HGB-driven case, the propagation dynamics of the 25$^{\text{th}}$ harmonic shown in Fig. \ref{fig:attosecond_pulses} a) closely follow those of the theoretical HGB presented in Fig.~\ref{fig:setup} a). 
As a result of the coherence of the HHG process, the spatial structure of the HGB driver (Fig.~\ref{fig:setup}) is transferred to the harmonics. Consequently, not just the ring shape is reproduced in the focal plane, but also a hot-spot before the focus, where the harmonic intensity is maximized. 
Remarkably, this hot-spot naturally concentrates the harmonic energy within a more confined spatial region compared to that of the Gaussian-driven case. Moreover, the variation of the harmonic hot-spot position along the propagation axis is substantially smaller. This effect is illustrated by the green dots in Fig.~\ref{fig:attosecond_pulses} b), which correspond to a standard HGB-driven deviation of $\rm{STD_{HGB}}\approx \SI{90}{\micro\meter}$, nearly half that of the Gaussian-driven case.
The break of symmetry in respect to the focal plane, i.e. the fact that there is no equally intense hot-spot on the other side, is related to the residual impact of the dipole phase.


The reduced deviation in the refocusing of all harmonic orders opens up the possibility to generate more intense attosecond pulse trains due to the improvement in the overlap of the different spectral components. This is illustrated in the simulation results presented in Fig.~\ref{fig:attosecond_pulses} c), where we show the spatio-temporal distribution of the harmonic radiation calculated at the position where the peak intensity position of the attosecond pulse train is obtained, which corresponds to the dashed lines in Fig.~\ref{fig:attosecond_pulses} b). The attosecond pulse train is obtained by filtering out the harmonics below the $23^\mathrm{rd}$ order. The top panel corresponds to the Gaussian-driven case, while the bottom one represents the HGB-driven case. Although the spatial extent of the resulting attosecond pulse trains is similar in both, the improved refocusing properties and higher spectral intensity in the HGB-driven case lead to significantly more intense attosecond pulses. \textcolor{black}{In both cases, the spatiotemporal distribution of each individual pulse show a slight wavefront curvature, which is attributed to chromatic aberrations due to the intrinsic dipole phase, as discussed in the Supplementary Information S2}. The on-axis temporal emission is shown in Fig.~\ref{fig:attosecond_pulses} d), for both the Gaussian-driven (blue) and HGB-driven (green) cases. Notably, the attosecond pulse intensity doubles when using the HGB driver. The temporal duration (FWHM) of the attosecond pulses remains similar ($\tau_{\rm{Gaussian}}\sim 210\:\rm{as}$ and $\tau_{\text{HGB}}\sim 175\:\rm{as}$).

Finally, we conducted a systematic numerical study of the attosecond pulse train peak intensity obtained when using higher-order HGBs, with $n\geq1$, maintaining the same peak driving field intensity. Fig.~\ref{fig:attosecond_pulses} e) shows the attosecond pulse peak intensity of $n$-fold HGB-driven, normalized to  Gaussian-driven ($n=0$) HHG. The red dot corresponds to the factor of $~2$ depicted in panel d). Our results predict that the attosecond pulse peak intensity increases with $n$, reaching a factor of $3$ compared to the Gaussian-driven case for $n=10$. As the HGB effective generation area increases with $n$, the total driving pulse energy employed to drive the up-conversion process scales as $(e/2n)^{2n}\Gamma(1+2n)$, where $\Gamma(\cdot)$ is the gamma function. 
Thereby, the use of higher order HGB drivers allows to deliver more energy into the gas jet without increasing the intensity, while maintaining the Rayleigh length constant for optimizing the physical footprint of the HHG setup.

\section{Conclusions}
Our results introduce a novel approach for generating spatially confined high-order harmonic beams and intense attosecond pulses. The use of HGBs as driving beams in HHG redistributes the laser pulse energy in a ring-shaped intensity profile with flat phase, allowing for greater energy deposition into the HHG process. 
Usually, to accommodate higher driving pulse energy, one would increase the focal length, which can result in long and bulky attosecond beamlines.
For HGB drivers, on the other hand, the focal length can be reduced, promoting more compact setups.
In addition, for HGBs, in contrast to Gaussian beams, the harmonic divergence decreases with increasing harmonic order.

Our numerical simulations confirm that HGB-driven HHG provides optimized refocusing conditions, with a small focus and little longitudinal spread over the whole harmonic spectrum. This leads to significantly more intense attosecond pulses compared to standard Gaussian-driven schemes. 
While $n=1$ HGBs provides a factor of two increase in intensity, the use of higher HGBs (where the driving ring size increases) further enhances the intensity, reaching a factor of three for $n=10$, along with a significant  increase in the APT energy. In contrast, a conventional beamline using a Gaussian beam with the same pulse energy would need to be at least three times longer to achieve similar performance.

Our study demonstrates that HGB-driven HHG is an ideal candidate for compact attosecond and high-order harmonic sources---even when driven by multi-terawatt laser systems---paving the way for their use in applications such as diffraction or nonlinear spectroscopy. 
This method provides a promising route towards compact intense EUV sources with excellent temporal coherence, achieving intensities  closer to those delivered by large-scale facilities as X-ray free electron lasers \cite{Franz2024, Driver2024}.


\section{Acknowledgments}

This project has received funding from the European Research Council (ERC) under the European Union’s Horizon 2020 research and innovation programme	(grant	agreements No 851201 and No. 871124). C.H.-G. and L.P. acknowledge funding from	Ministerio de	Ciencia	e Innovacio\'on	(Grant PID2022-142340NB-I00) and the Department of Education of the Junta de Castilla y Le\'on and FEDER Funds (Escalera de Excelencia CLU-2023-1-02).The authors acknowledge support from the Swedish Research Council (Grant Nos. 2013-8185, 2021-04691, 2022-03519, and 2023-04603), the European Research Council (advanced grant QPAP, Grant No. 884900), the Crafoord Foundation, and the Knut and Alice Wallenberg Foundation. A.L. acknowledges support from the Knut and Alice Wallenberg Foundation through the Wallenberg Centre for Quantum Technology (WACQT). We acknowledge the computer resources at MareNostrum and the technical support provided by the Barcelona Supercomputing Center (FI-2024-2-0010 and FI-2024-3-0035).